\documentstyle[12pt]{article}
\newcommand{\nc}{\newcommand}
\nc{\renc}{\renewcommand}

\nc{\half}{{\textstyle{1\over2}}}

\nc{\etal}{\mbox{\it et al. }}
\nc{\ie}{{\it i.e.}}
\nc{\eg}{{\it e.g.}}

\renc{\thefootnote}{\arabic{footnote}}
\nc{\capt}[1]{{\bf Figure.} {\small\sl #1}}

\nc{\eqs}[2]{\mbox{Eqs.~(\ref{#1},\,\ref{#2})}}
\nc{\eq}[1]{\mbox{Eq.~(\ref{#1})}}

\nc{\figs}[2]{\mbox{Figs.~(\ref{#1},\,\ref{#2})}}
\nc{\fig}[1]{\mbox{Fig~.(\ref{#1})}}

\nc{\tag}[1]{\label{#1} \marginpar{{\footnotesize #1}}}
\nc{\mtag}[1]{\label{#1} \mbox{\marginpar{{\footnotesize #1}}}}
\renc{\baselinestretch}{1.2}
\newlength{\overeqskip}
\newlength{\undereqskip}
\nc{\be}[1]{\begin{equation} \mbox{$\label{#1}$}}
\nc{\bea}[1]{\begin{eqnarray} \mbox{$\label{#1}$}}
\nc{\Section}[2]{\section{#2}\label{#1}}
\nc{\Bibitem}[1]{\bibitem{#1}}
\nc{\Label}[1]{\label{#1}}

\nc{\eea}{\vspace{\undereqskip}\end{eqnarray}}
\nc{\ee}{\vspace{\undereqskip}\end{equation}}
\nc{\bdm}{\begin{displaymath}}
\nc{\edm}{\end{displaymath}}
\nc{\dpsty}{\displaystyle}
\nc{\bc}{\begin{center}}
\nc{\ec}{\end{center}}
\nc{\ba}{\begin{array}}
\nc{\ea}{\end{array}}
\nc{\bab}{\begin{abstract}}
\nc{\eab}{\end{abstract}}
\nc{\btab}{\begin{tabular}}
\nc{\etab}{\end{tabular}}
\nc{\bit}{\begin{itemize}}
\nc{\eit}{\end{itemize}}
\nc{\ben}{\begin{enumerate}}
\nc{\een}{\end{enumerate}}
\nc{\bfig}{\begin{figure}}
\nc{\efig}{\end{figure}}
\nc{\arreq}{&\!=\!&}
\nc{\arrmi}{&\!-\!&}
\nc{\arrpl}{&\!+\!&}
\nc{\arrap}{&\!\!\!\approx\!\!\!&}
\nc{\non}{\nonumber\\*}
\nc{\align}{\!\!\!\!\!\!\!\!&&}

\def\lsim{\; \raise0.3ex\hbox{$<$\kern-0.75em
      \raise-1.1ex\hbox{$\sim$}}\; }
\def\gsim{\; \raise0.3ex\hbox{$>$\kern-0.75em
      \raise-1.1ex\hbox{$\sim$}}\; }
\nc{\DOT}{\hspace{-0.08in}{\bf .}\hspace{0.1in}}
\nc{\Laada}{\hbox {$\sqcap$ \kern -1em $\sqcup$}}
\nc\loota{{\scriptstyle\sqcap\kern-0.55em\hbox{$\scriptstyle\sqcup$}}}
\nc\Loota{{\sqcap\kern-0.65em\hbox{$\sqcup$}}}
\nc\laada{\Loota}
\nc{\qed}{\hskip 3em \hbox{\BOX} \vskip 2ex}

\nc{\real}{{\rm I \! R}}
\nc{\Z}{{\sf Z \!\!\! Z}}
\nc{\complex}{{\rm C\!\!\! {\sf I}\,\,}}
\def\bigid{\leavevmode\hbox{\small1\kern-3.8pt\normalsize1}}
\def\id{\leavevmode\hbox{\small1\kern-3.3pt\normalsize1}}
\nc{\slask}{\!\!\!/}
\nc{\bis}{{\prime\prime}}
\nc{\pa}{\partial}
\nc{\na}{\nabla}
\nc{\ra}{\rangle}
\nc{\la}{\langle}
\nc{\goto}{\rightarrow}
\nc{\swap}{\leftrightarrow}

\nc{\EE}[1]{ \mbox{$\cdot10^{#1}$} }
\nc{\abs}[1]{\left|#1\right|}
\nc{\at}[2]{\left.#1\right|_{#2}}
\nc{\norm}[1]{\|#1\|}
\nc{\abscut}[2]{\Abs{#1}_{\scriptscriptstyle#2}}
\nc{\vek}[1]{{\rm\bf #1}}
\nc{\integral}[2]{\int\limits_{#1}^{#2}}
\nc{\inv}[1]{\frac{1}{#1}}
\nc{\dd}[2]{{{\partial #1}\over{\partial #2}}}
\nc{\ddd}[2]{{{{\partial}^2 #1}\over{\partial {#2}^2}}}
\nc{\dddd}[3]{{{{\partial}^2 #1}\over
        {\partial #2 \partial #3}}}
\nc{\dder}[2]{{{d #1}\over{d #2}}}
\nc{\ddder}[2]{{{d^2 #1}\over{d {#2}^2}}}
\nc{\dddder}[3]{{d^2 #1}\over
        {d #2 d #3}}
\nc{\dx}[1]{d\,^{#1}x}
\nc{\dy}[1]{d\,^{#1}y}
\nc{\dz}[1]{d\,^{#1}z}
\nc{\dl}[1]{\frac{d\,^{#1}l}{(2\pi)^{#1}}}
\nc{\dk}[1]{\frac{d\,^{#1}k}{(2\pi)^{#1}}}
\nc{\dq}[1]{\frac{d\,^{#1}q}{(2\pi)^{#1}}}

\nc{\cc}{\mbox{$c.c.$ }}
\nc{\hc}{\mbox{$h.c.$ }}
\nc{\cf}{cf.\ }
\nc{\erfc}{{\rm erfc}}
\nc{\Tr}{{\rm Tr\,}}
\nc{\tr}{{\rm tr\,}}
\nc{\pol}{{\rm pol}}
\nc{\sign}{{\rm sign}}
\nc{\bfT}{{\bf T }}

\def\MeV{{\rm\ MeV}}

\nc{\cA}{{\cal A}}
\nc{\cB}{{\cal B}}
\nc{\cD}{{\cal D}}
\nc{\cE}{{\cal E}}
\nc{\cG}{{\cal G}}
\nc{\cH}{{\cal H}}
\nc{\cL}{{\cal L}}
\nc{\cO}{{\cal O}}
\nc{\cT}{{\cal T}}
\nc{\cN}{{\cal N}}
\nc{\rvac}[1]{|{\cal O}#1\rangle}
\nc{\lvac}[1]{\langle{\cal O}#1|}
\nc{\rvacb}[1]{|{\cal O}_\beta #1\rangle}
\nc{\lvacb}[1]{\langle{\cal O}_\beta #1 |}
\nc{\bb}{\bar{\beta}}
\nc{\bt}{\tilde{\beta}}
\nc{\ctH}{\tilde{\cal H}}
\nc{\chH}{\hat{\cal H}}
\nc{\1}{\aa}
\nc{\2}{\"{a}}
\nc{\3}{\"{o}}
\nc{\4}{\AA}
\nc{\5}{\"{A}}
\nc{\6}{\"{O}}
\nc{\al}{\alpha}
\nc{\g}{\gamma}
\nc{\Del}{\Delta}
\nc{\e}{\epsilon}
\nc{\eps}{\epsilon}
\nc{\lam}{\lambda}
\nc{\om}{\omega}
\nc{\Om}{\Omega}
\nc{\ve}{\varepsilon}
\nc{\mn}{{\mu\nu}}
\nc{\k}{\kappa}
\nc{\vp}{\varphi}

\nc{\advp}[3]{{\it  Adv.\ in\ Phys.\ }{{\bf #1} {(#2)} {#3}}}
\nc{\annp}[3]{{\it  Ann.\ Phys.\ (N.Y.)\ }{{\bf #1} {(#2)} {#3}}}
\nc{\apl}[3]{{\it  Appl. Phys. Lett. }{{\bf #1} {(#2)} {#3}}}
\nc{\apj}[3]{{\it  Ap.\ J.\ }{{\bf #1} {(#2)} {#3}}}
\nc{\apjl}[3]{{\it  Ap.\ J.\ Lett.\ }{{\bf #1} {(#2)} {#3}}}
\nc{\app}[3]{{\it Astropart.\ Phys.\ }{{\bf #1} {(#2)} {#3}}}
\nc{\cmp}[3]{{\it  Comm.\ Math.\ Phys.\ }{{ \bf #1} {(#2)} {#3}}}
\nc{\cqg}[3]{{\it  Class.\ Quant.\ Grav.\ }{{\bf #1} {(#2)} {#3}}}
\nc{\epl}[3]{{\it  Europhys.\ Lett.\ }{{\bf #1} {(#2)} {#3}}}
\nc{\ijmp}[3]{{\it Int.\ J.\ Mod.\ Phys.\ }{{\bf #1} {(#2)} {#3}}}
\nc{\ijtp}[3]{{\it Int.\ J.\ Theor.\ Phys.\ }{{\bf #1} {(#2)} {#3}}}
\nc{\jmp}[3]{{\it  J.\ Math.\ Phys.\ }{{ \bf #1} {(#2)} {#3}}}
\nc{\jpa}[3]{{\it  J.\ Phys.\ A\ }{{\bf #1} {(#2)} {#3}}}
\nc{\jpc}[3]{{\it  J.\ Phys.\ C\ }{{\bf #1} {(#2)} {#3}}}
\nc{\jap}[3]{{\it J.\ Appl.\ Phys.\ }{{\bf #1} {(#2)} {#3}}}
\nc{\jpsj}[3]{{\it J.\ Phys.\ Soc.\ Japan\ }{{\bf #1} {(#2)} {#3}}}
\nc{\lmp}[3]{{\it Lett.\ Math.\ Phys.\ }{{\bf #1} {(#2)} {#3}}}
\nc{\mpl}[3]{{\it  Mod.\ Phys.\ Lett.\ }{{\bf #1} {(#2)} {#3}}}
\nc{\ncim}[3]{{\it  Nuov.\ Cim.\ }{{\bf #1} {(#2)} {#3}}}
\nc{\np}[3]{{\it  Nucl.\ Phys.\ }{{\bf #1} {(#2)} {#3}}}
\nc{\pr}[3]{{\it Phys.\ Rev.\ }{{\bf #1} {(#2)} {#3}}}
\nc{\pra}[3]{{\it  Phys.\ Rev.\ A\ }{{\bf #1} {(#2)} {#3}}}
\nc{\prb}[3]{{\it  Phys.\ Rev.\ B\ }{{{\bf #1} {(#2)} {#3}}}}
\nc{\prc}[3]{{\it  Phys.\ Rev.\ C\ }{{\bf #1} {(#2)} {#3}}}
\nc{\prd}[3]{{\it  Phys.\ Rev.\ D\ }{{\bf #1} {(#2)} {#3}}}
\nc{\prl}[3]{{\it Phys.\ Rev.\ Lett.\ }{{\bf #1} {(#2)} {#3}}}
\nc{\pl}[3]{{\it  Phys.\ Lett.\ }{{\bf #1} {(#2)} {#3}}}
\nc{\prep}[3]{{\it Phys\. Rep.\ }{{\bf #1} {(#2)} {#3}}}
\nc{\prsl}[3]{{\it Proc.\ R.\ Soc.\ London\ }{{\bf #1} {(#2)} {#3}}}
\nc{\ptp}[3]{{\it  Prog.\ Theor.\ Phys.\ }{{\bf #1} {(#2)} {#3}}}
\nc{\ptps}[3]{{\it  Prog\ Theor.\ Phys.\ suppl.\ }{{\bf #1} {(#2)} {#3}}}
\nc{\physa}[3]{{\it  Physica\ A\ }{{\bf #1} {(#2)} {#3}}}
\nc{\physb}[3]{{\it  Physica\ B\ }{{\bf #1} {(#2)} {#3}}}
\nc{\phys}[3]{{\it Physica\ }{{\bf #1} {(#2)} {#3}}}
\nc{\rmp}[3]{{\it  Rev.\ Mod.\ Phys.\ }{{\bf #1} {(#2)} {#3}}}
\nc{\rpp}[3]{{\it Rep.\ Prog.\ Phys.\ }{{\bf #1} {(#2)} {#3}}}
\nc{\sjnp}[3]{{\it Sov.\ J.\ Nucl.\ Phys.\ }{{\bf #1} {(#2)} {#3}}}
\nc{\spjetp}[3]{{\it Sov.\ Phys.\ JETP\ }{{\bf #1} {(#2)} {#3}}}
\nc{\yf}[3]{{\it Yad.\ Fiz.\ }{{\bf #1} {(#2)} {#3}}}
\nc{\zetp}[3]{{\it Zh.\ Eksp.\ Teor.\ Fiz.\  }{{\bf #1}  {(#2)} {#3}}}
\nc{\zp}[3]{{\it Z.\ Phys.\ }{{\bf #1} {(#2)} {#3}}}
\nc{\ibid}[3]{{\sl ibid.\ }{{\bf #1} {#2} {#3}}}
\nc{\rf}[1]{(\ref{#1})}
\nc{\nn}{\nonumber \\*}
\nc{\bfB}{\bf{B}}
\nc{\bfv}{\bf{v}}
\nc{\bfx}{\bf{x}}
\nc{\bfy}{\bf{y}}
\nc{\vx}{\vec{x}}
\nc{\vy}{\vec{y}}
\nc{\oB}{\overline{B}}
\nc{\oI}{\overline{I}}
\nc{\oR}{\overline{R}}
\nc{\rar}{\rightarrow}
\nc{\ti}{\times}
\nc{\slsh}{\hskip-5pt/}
\nc{\sm}{Standard~Model~}
\nc{\MP}{M_{\rm Pl}}
\nc{\tp}{t_{\rm Pl}}
\nc{\ave}{\bar{E}}

\nc{\eff}{{\rm eff}}
\nc{\kk}{\vek{k}}
\nc{\pp}{{\rm p}}
\nc{\ga}{g_{a\gamma}}
\nc{\vv}{\\}
\nc{\eee}{{\bf E}}
\nc{\bbb}{{\bf B}}
\nc{\qcd}{T_{\rm QCD}}
\nc{\G}{\rm \ G}
\def\vec#1{{\bf #1}}
\begin{document}

{\title{\vskip-2truecm{\hfill {{\small HU-TFT-96-10\\
        }}\vskip 1truecm}
{\bf Electrical conductivity in the early universe}}


{\author{
{\sc Jarkko Ahonen$^{1}$  }\\
{\sl\small Research Institute for Theoretical Physics, P.O. Box 9,}\\
{\sl\small FIN-00014 University of Helsinki,
Finland} \\
and \\
{\sc Kari Enqvist$^{2}$}\\
{\sl\small Department of Physics, P.O. Box 9,
FIN-00014 University of Helsinki,
Finland}
}
\maketitle
\vspace{2cm}
\begin{abstract}
\noindent
We solve numerically the Boltzmann equation in the early universe in the
presence of a constant electric field and find the electrical conductivity
$\sigma$ in the range  $1\MeV\lsim T\lsim M_W$. The main contribution to 
$\sigma$ is
shown to be due to leptonic interactions. For $T\lsim 100\MeV$ 
we find $\sigma\simeq 0.76T$ while at $T\simeq M_W$ we obtain
$\sigma\simeq 6.7T$ 
\end{abstract}
\vfil
\footnoterule
{\small $^1$jtahonen@rock.helsinki.fi};
{\small $^2$enqvist@pcu.helsinki.fi};
\thispagestyle{empty}
\newpage
\setcounter{page}{1}


Recently, there has been much interest in primordial magnetic fields.
Primordial  fields might act as the seed field for the dynamo mechanism 
which may be  responsible for
the observed galactic magnetic fields \cite{fields}.
These have been measured both in the
Milky Way and in other spiral galaxies, including their halos (which perhaps
can be viewed as an indication of the primordial origin of the seed field). 
Typically the
strength of the observed  field is of the order of ${\cal O}(10^{-6})$ G. 

It has been suggested 
that primordial magnetic fields might be generated as a consequence of 
the early cosmic phase
transitions, such as the electroweak or the QCD phase transitions \cite{pt}. 
These would then be small scale random fields.
They would  be imprinted on the comoving plasma and dissipate very slowly 
\cite{dissip} because the plasma  of the early universe  
is known to have a very large electrical conductivity \cite{high}.
Indeed, on dimensional grounds alone one can argue that conductivity scales as 
$\sigma\sim T/\alpha$, where $\alpha$ is the coupling constant squared
associated with the scattering processes of the charged particles
in the plasma. 
As a consequence, the magnetic Reynolds number of the universe, given by
$R\sim v\sigma H^{-1}$ where $H\equiv \dot R/R\sim T^2/M_{\rm Pl}$ is the Hubble parameter
and $v$ the bulk velocity,
is huge. Despite its smallness, diffusion can play an important role in
the subsequent evolution of the microscopic random magnetic field. 
The main issue here is just how the small scale field, the origin of which is
microphysics,  gets amplified to a large scale seed field.
In a recent paper Baym, B\"odeker and McLerran \cite{larry} have suggested that
in a first order electroweak phase transition, turbulence generated by
the colliding shock fronts of the phase transition bubbles could give rise
to large fields. More generally, a recent numerical simulation \cite{beo} 
of the
full magnetohydronamical equations in an expanding universe shows clearly
a transfer of energy from small scales to large scales, albeit in 2 dimensions.
Because of the very large Reynolds number, real numerical simulations are 
not presently realistic. However, in 1+3 dimensions one may apply 
the so-called shell (or cascade) model, which has proved very succesful
in studies of  hydrodynamical turbulence, to the early universe, with the
result that the scale of the magnetic field fluctuations indeed increases
very rapidly \cite{beo}.

The previous estimates for the conductivity
$\sigma$ have, however, been rough 
order of magnitude estimates only. The proper
way to derive electrical conductivity of the early universe is to
consider plasma in an electric field, and then use the Boltzmann equations
to find the bulk motion, which  serves to define conductivity through Ohm's 
law. To be able to do this, one  first has to compute the collision integral,
which is the weighted sum of the matrix elements of all the relevant
processes leading to microscopic diffusion of the bulk motion. In the
present paper we compute the collision integrals for leptons and hadrons
(quarks) separately and solve the Boltzmann equations in the approximation
where bulk velocity is much smaller than the thermal velocity of the plasma
particles. This means that we treat the effect of the electric field as
a perturbation in an otherwise homogeneous and isotropic background.

The Boltzmann equation for the  distribution function 
$f(x(\lambda ),p(\lambda ))$ of a charged particle 
in an electromagnetic field is found by requiring that along the world line,
para\-metriz\-ed by $\lambda$,
the total change in $f$ equals the collision integral. Using
the equation of motion of a point particle with electric charge $q$, which 
in curved space reads
\be{einstein}
\frac{dp^\mu}{d\lambda}=-\Gamma^{\mu}_{\alpha\beta}p^{\alpha}p^{\beta}
+qF^{\mu}_{\beta}p^{\beta}~,
\ee
one  can cast the Boltzmann equation in the form
\be{boltzmann}
\frac{\partial f}{\partial t}p^{0}+
\vec{p}\cdot\nabla f-\frac{\partial f}{\partial p^0}p^2
R\dot{R}-2\frac{\dot{R}}{R}p^0p^i\frac{\partial f}{\partial
p^i}-q\frac{\partial f}{\partial p^0}\vec{E}\cdot\vec{p}-q\frac{\partial f}
{\partial p^i}p^0E^i=C(p,t)p^0~,
\ee
where $C(p,t)$ is the collision integral.
Here we have assumed a 
Robertson-Walker metric with a scale factor $R$
for the background, and
we have defined $F_i^0=-E_i$ and $F_i^j=\epsilon^{ijk}B_k$.
We also prefer to use co-moving coordinates and define
\be{co-mo}
\tilde{f}(p,t)=\int\delta
(p_{0}-(\vec{p}^2R^2+m^2)^{1/2})f(\vec{p},p_{0},t) dp_{0}~.
\ee
Inserting this into \eq{boltzmann} and integrating we find,
making use of the local momentum defined as $\tilde\vec p=R\vec p$,
\be{perusyht}
\frac{\partial \tilde{f}}{\partial t}-\frac{\dot{R}}{R}\tilde p^i
\frac{\partial\tilde{f}}{\partial
\tilde p^i}-{q}\frac{\partial\tilde{f}}
{\partial \tilde p^i}E^i=C(\tilde p,t)~.
\ee
Because of the assumed isotropy of the background,   the
terms $\vec{p}\cdot\nabla f$ and $\vec{E}\cdot \vec{p}{\partial f}/{\partial
p_{0}}$ have been dropped. This is justified if one 
assumes, as we do, that the effect of the electromagnetic field on the
distributions can be considered as a perturbation. From now on, we also
drop the tildes for brevity.

The collision integral encompasses all the scattering processes relevant
for the diffusion of the charged test particle. For electromagnetism the
leading processes
are $2\to 2$ reactions (we do not consider thermal effects such as the
plasmon decay). Thus
\be{collint}
C(p,t)=\frac{1}{p_0}\int dP_{b}dP_{c}dP_{d}(2\pi )^4
\delta^4 (p+p_b-p_c-p_d)|M(Ab\to cd)|^2{\cal F},
\ee
where $p$ is the four-momentum of the charged test particle  $A$, and
we have defined
$dP_{i}\equiv {d^3 \vec p_i}/({(2\pi )^3}{2E(p_i)})$. The factor
${\cal F}$ contains the distribution functions of the initial and final
states. For fermions one must include the Pauli blocking factors and for
bosons the enhancement factors. Below are two examples of the 
${\cal F}$-factor. The
first is for the case when all the particles in the reaction 
are fermions, and in the second
case particles $b$ and $c$ are bosons, such as is the case for 
Compton scattering:
\be{fermi}
{\cal F}=\cases{{
[1-f(p)][1-f_b(p_b)]f_c(p_c)f_d(p_d)-[1-f_c(p_c)][1-f_d(p_d)]
f(p)f_b(p_b)~,}\cr
{[1-f(p)][1+f_b(p_b)]f_c(p_c)f_d(p_d)-
[1-f_d(p_d)][1+f_c(p_c)]f(p)f_b(p_b)~.}\cr}
\ee
Above the time dependence is not shown explicitly.

Let us now assume that there exists a constant electric field in the
early universe\footnote{Of course, we do not claim such a field actually exists
but rather use it as a probe of the plasma properties.}
(constant in the sense that its coherence length is 
larger than the mean free path of charged particles) and treat its effect  
as a perturbation on the distribution of the test particle. We write
$f=f_{0}+\delta f$, where $f_{0}$ is
the equilibrium
distribution and $\delta f$ the small perturbation. Inserting this into
the collision integral \eq{collint} results in 

\bea{pertcoll}
C(p,t)&=&-\frac{1}{p_0}\int dP_{b}dP_{c}dP_{d}(2\pi )^4\delta^4 (p + p_b -p_c -p_d)\cr
&\times&|M|^2\big([f_{0c}f_{0d}[1-f_{0b}]+f_{0b}[1-f_{0c}][1-f_{0d}]\big)
\delta f~
\equiv\hat C(p)\delta f(p,t).
\eea

\eq{pertcoll} assumes that all the particles in the reaction are fermions;
generalization to other cases is straightforward.

We have assumed that all the particles b,c and d have an equilibrium 
distribution. Therefore, if also 
the test particle has an equilibrium distribution,
the collision integral vanishes. Thus only the
perturbation term in \eq{pertcoll} survives. 
Note that  the equilibrium distribution depends 
only on the momentum. 

Treating the electric field
as a small perturbation, the Boltzmann equation can be
linearized and reads
\be{linboltz}
\frac{\partial \delta f}{\partial t}-\frac{\dot{R}}{R}p^i\frac{\partial
\delta f}{\partial
p^i} -{q}\frac{\partial f}{\partial p^i}E^i=\hat C(p)\delta f(p,t)
\ee
with $\hat C(p)$ defined in \eq{pertcoll}. Finally, assuming that 
 $E^i\gg H\vert\vec p\vert\equiv\vert\vec p\vert$, 
we may search for stationary perturbations,
for which one easily obtains
\be{delta}
\delta f(p)=-\frac{q}{\hat C(p)}\frac{\partial f_0(p)}{\partial p^i}E^i~.
\ee

The induced current density is given by
\be{jii}
\vec j=q\int\delta f(p)\vec{p}d^3\vec{p}~,
\ee
and conductivity $\sigma_A$, associated with a given particle species $A$, 
is defined by
\be{cond}
\vec j_A={\sigma_A}\vec E~.
\ee
Thus the contribution of a single species $A$ to conductivity is
$\sigma_A\sim 1/\sum\vert M(AX\to Y)\vert^2$, where the sum is over
all the processes which scatter the test particle $A$. For the purpose
of conductivity, we may view the mixture of different particle species, such
as is found in the early universe, a multicomponent fluid. The flow of each
component contributes to the total current and adds up to the total 
conductivity, which reads
\be{condtot}
\sigma_{\rm tot}=\sum_A\sigma_A~,
\ee
where the sum is over all the relativistic charged species present in
the thermal bath. Note that the
total conductivity is dominated by the species that has the weakest
interaction. This is because the weaker the interaction, the longer time
the current flow  is maintained.

We consider the temperature interval $1\MeV \lsim T\lsim M_W$, and
make the simple, crude assumption that particles appear in the thermal
bath only when temperature is greater than their mass. Thus below
$100$ MeV, for example, the only charged particles present are the
electrons and positrons.
When $T\gsim \qcd$, also the quarks should be counted in. Their
main interactions are strong, so that their electromagnetic interactions
may be neglected. The list of the relevant reactions for the leptons is presented
in  Table 1, and in  Table 2 the same for the quarks, together with the matrix 
element squared (modulo factors of $\pi$ and $\alpha$). A technical
point is that for the t-channel reactions there arises an infrared
singularity, which is known to be regulated by thermal effects.
These we approximate by giving fermions, photons and gluons a Debye mass
in the t-channel and u-channel propagator; otherwise all fermions are 
assumed to be massless 
(consistency requires that the external particle masses are kept non-thermal).
The thermal masses are given by \cite{thermal}

\bea{masses}
m_l^2&=&\frac{e^2}{8}T^2\simeq 0.0115 T^2,\cr
m_q^2&=&\frac{g_{s}^{2}}{6}T^2\simeq 0.251 T^2 ,\cr
m_\gamma^2&=&\frac{e^2}{3}(\Sigma_{l}Q_{l}^2+3\Sigma_{q}Q_{q}^{2})T^2\simeq
0.0306(\Sigma_{l}Q_{l}^2+3\Sigma_{q}Q_{q}^{2})T^2,\cr
m_g^2&=&g_{s}^{2}(3+\frac{N_{f}}{3}) T^2 \simeq 
1.508(3+\frac{N_{f}}{3})T^2.
\eea
where $N_{f}$ is the number of quark families present, the sums are over all particles with
$m\le T$, and we adopt the values $g_{s}^{2}=4\pi\alpha_s\simeq 1.508$
and $e^2=4\pi\alpha_{e}\simeq 0.0917$, and $\qcd=200 \MeV$.

{\begin{table*}
\centering
\caption[t2]{Leptonic processes for $\mbox{e}^{-}$, and their matrix
elements squared}
\begin{tabular}{|c|c|}

$\mbox{$e$}^{-}\mbox{$e$}^{-}\rightarrow\gamma\gamma$     &  
$\frac{u^2+t^2}{ut}$ \\
\hline
$\mbox{$e$}^{-}\gamma\rightarrow\mbox{$e$}^{-}\gamma$ & $\frac{u^2+s^2}{-us}$ \\
\hline
$\mbox{$e$}^{-}\mbox{$e$}^{+}\rightarrow\mbox{$e$}^{-}\mbox{$e$}^{+}$ &
$\frac{s^4+t^4+u^2(s+t)^2}{s^2t^2}$ \\
\hline
$\mbox{$e$}^{-}\mbox{$e$}^{-}\rightarrow\mbox{$e$}^{-}\mbox{$e$}^{-}$ &
$\frac{u^4+t^4+s^2(u+t)^2}{u^2t^2}$ \\
\hline
$\mbox{$e$}^{-}\mbox{$l$}^{-}\rightarrow\mbox{$e$}^{-}\mbox{$l$}^{-}$ &
$\frac{s^2+u^2}{t^2}$   \\
\hline
$\mbox{$e$}^{-}\mbox{$e$}^{+}\rightarrow\mbox{$l$}^{-}\mbox{$l$}^{+}$ &
$\frac{u^2+t^2}{s^2}$ \\
\hline
$\mbox{$e$}^{-}\mbox{$e$}^{+}\rightarrow q\bar q$ &
$\frac{u^2+t^2}{s^2}$
\end{tabular}
\label{t2}
\end{table*}
\null
\begin{table*}
\centering
\caption[t3]{QCD-processes, and their matrix
elements squared}
\begin{tabular}{|c|c|}
$\mbox{$q$}\bar{\mbox{$q$}}\rightarrow gg$  &
$\frac{(u^2+t^2)(4u^2+4t^2-ut)}{9s^2ut}$ \\
\hline
$\mbox{$q$}g\rightarrow\mbox{$q$}g$ &
$-\frac{(u^2+s^2)(4u^2+4s^2-us)}{9t^2us}$ \\
\hline
$\mbox{$q$}\bar{\mbox{$q$}}\rightarrow\mbox{$q$}\bar{\mbox{$q$}}$ &
$\frac{s^4+t^4+u^2(s+t)^2}{s^2t^2}$ \\
\hline
$\mbox{$q$}\mbox{$q$}\rightarrow\mbox{$q$}\mbox{$q$}$ &
$\frac{u^4+t^4+s^2(u+t)^2}{u^2t^2}$ \\
\hline
$\mbox{$q$}\mbox{$q$}'\rightarrow\mbox{$q$}\mbox{$q$}'$ &
$\frac{s^2+u^2}{t^2}$   \\
\hline
$\mbox{$q$}\bar{\mbox{$q$}}\rightarrow\mbox{$q'$}\bar{\mbox{$q$}}'$ &
$\frac{u^2+t^2}{s^2}$

\end{tabular}
\label{t3}
\end{table*}}
\vskip30pt

We have computed the perturbation $\delta f(p)$ numerically
by evaluating the collision integral by a simple Monte Carlo simulation. 
The result
agrees with the expected naive scaling law $\sigma \sim T$.
The form of $\delta f(p)$
is very similar for both leptons and quarks. As an example, in Fig. 1 we 
show $dj/dp=qp^3\delta f(p)$ for the electron at $T\lsim m_\mu$, where the peak at
$p \simeq 3T$ is evident. In Fig. 1 we also compare $\delta f(p)$ with the
equilibrium distribution $f_0(p)$ to demonstrate the region of validity
of our calculation, where we have chosen $E=10^{-3}T^2$ for definitess.
For most part, $\delta f(p)/f_0$ is a constant, except 
when $p\to 0$ where $\delta f(p)$ vanishes. This behaviour is due to
Compton scattering (and at higher temperatures also $l^+l^-$ annihilation),
for which the integrated matrix element squared gives rise to
$p$-depencence slower than the one inherent in $\hat C(p)$.
We should also point out that
our result is  sensitive to the the infrared cut-off provided
by the thermal masses of the internal propagators, given by \eq{masses},
and hence not strictly valid at very small $p$.
 This is because it is the t-channel
reactions which give the dominant contribution to the collision integral 
\eq{pertcoll}, and a straightforward numerical calculation  gives an
inversely linear dependence on the regulator in $\hat C (p)$.

With this caveat, the total leptonic conductivity,
together with the total hadronic conductivity, is shown in Fig. 2
for the range $1\MeV\lsim T\lsim M_W$, above which one would have
to account for the charged W's. However, purely electromagnetic
processes continue to dominate conductivity also well beyond the
Z-pole. The reader may compare the result presented in Fig. 2
with the text-book estimate \cite{high}
for relativistic electron gas scattering off heavy ions, which yields
$\sigma\simeq 14T$. The steps in the leptonic $\sigma$ reflect both the
appearance of new leptons in the thermal bath, as well as change in
the regulator (thermal photon mass) due to the appearance of new quarks 
and leptons.

\begin{figure}
\leavevmode
\centering
\vspace*{60mm}
\includegraphics{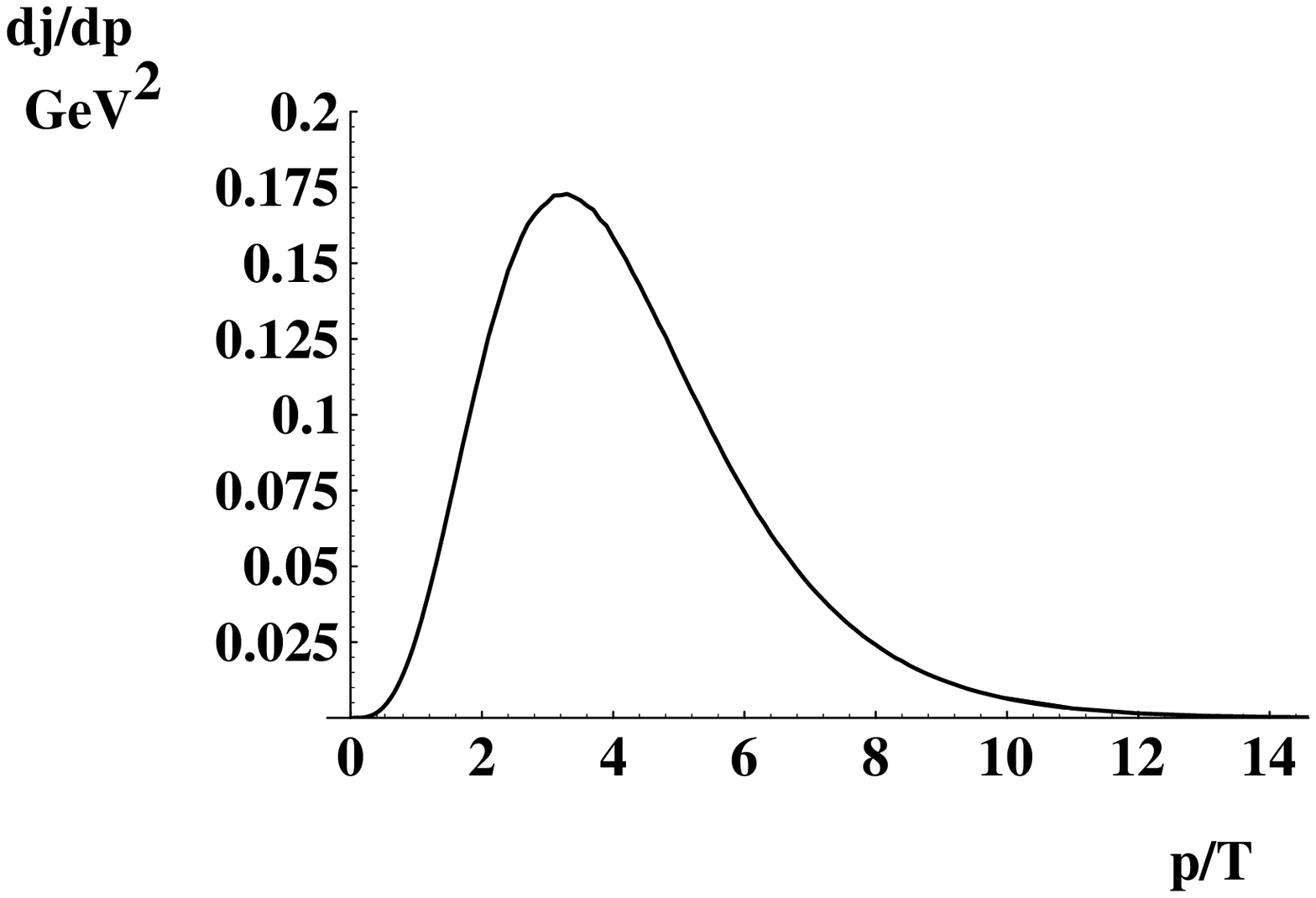}
\includegraphics{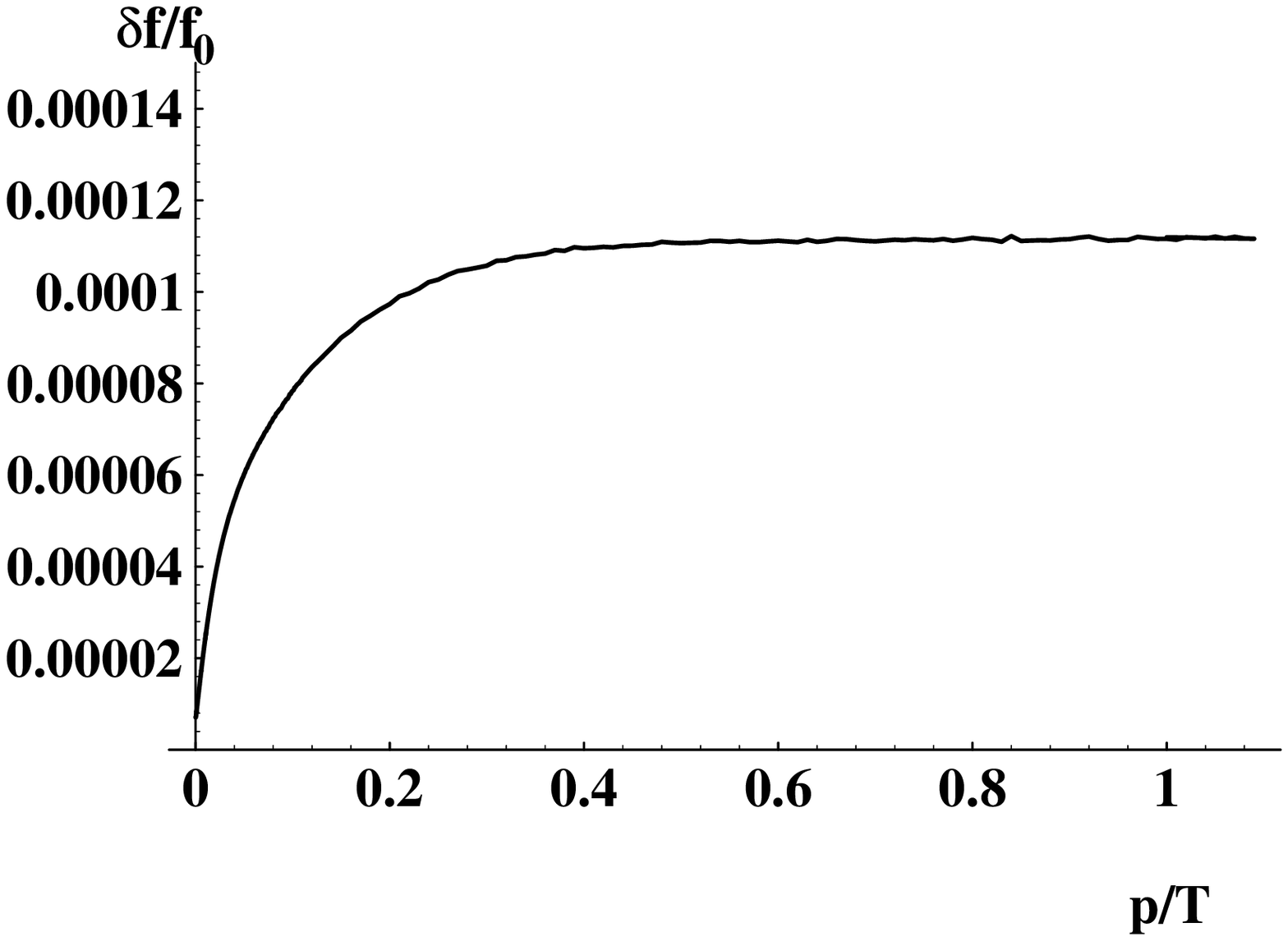}
\caption{$dj/dp$ and $\delta f /f_{0}$ for the electron at $T\lsim m_\mu$.
Here $E=10^{-3}T^2$.}
\label{giku}
\end{figure}

Introducing a thermal mass into the propagator is only
an approximation of the full thermal screening, which includes also
initial state and vertex corrections. Such corrections should remove the 
spurious infrared singularities in $\hat C(p)$. Our approximation should, 
however, be sufficient to produce a reliable order-of-magnitude
estimate of the electrical
conductivity in the very early universe. 
\begin{figure}
\leavevmode
\centering
\vspace*{60mm}
\includegraphics{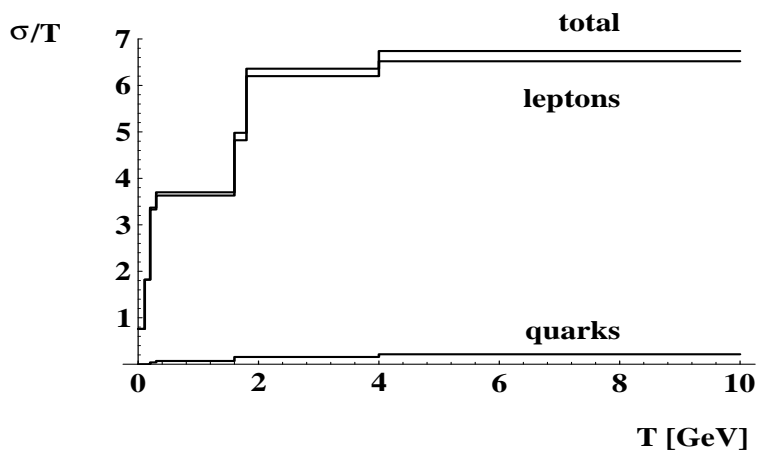}
\caption{$\sigma /T$ as a function of temperature.}
\label{kuvadmpw1}
\end{figure}

\vskip15pt
{\large\bf Acknowledgements}
\vskip5pt\noindent

J.A. wishes to thank Martti Havukainen and Hannes Uibo 
for illuminating discussions on
numerical algorithms.

\newpage


\end{document}